\documentclass[twocolumn,floats,showpacs,superscriptaddress,pre]{revtex4}
\usepackage{amsmath,amssymb,bm}
\usepackage{graphics}
\usepackage{epsfig}
\usepackage{graphicx}

\begin{document}

\title{CPT SYMMETRY IN THE MIRROR UNIVERSE}
\author{Natalia Gorobey}
\affiliation{Peter the Great Saint Petersburg Polytechnic University, Polytekhnicheskaya
29, 195251, St. Petersburg, Russia}
\author{Alexander Lukyanenko}
\email{alex.lukyan@mail.ru}
\affiliation{Peter the Great Saint Petersburg Polytechnic University, Polytekhnicheskaya
29, 195251, St. Petersburg, Russia}
\author{A. V. Goltsev}
\affiliation{Ioffe Physical- Technical Institute, Polytekhnicheskaya 26, 195251, St.
Petersburg, Russia}

\begin{abstract}
A model of a two-sheeted universe in the quantum theory of gravity is
proposed, based on the definition of 3D invariant and gauge-invariant proper
time of the universe. A uniform time in a closed universe is introduced in
the class of equivalent reference systems defined by the spectrum of a
Hermitian 3D operator on the space of Dirac bi-spinors, the equality to zero
of which is equivalent to a system of gravitational constraints. Based on
the analogy with the Lorentz-invariant quantum field theory in Minkowski
space and the definition of discrete C-, P- and T-transformations
separately, the principle of CPT symmetry in a two-sheeted universe is
formulated. A covariant quantum theory of matter gauge fields consistent
with the charge conjugation operation in a two-sheeted universe is proposed.
The CPT symmetric state of a homogeneous isotropic two-sheeted model of the
universe is constructed by an iterative method in which the no-boundary
Hartle--Hawking wave function is used as a zeroth approximation.
\end{abstract}

\maketitle







\section{\textbf{INTRODUCTION}}

As early as 1967, A.D. Sakharov \cite{Sah1} proposed the hypothesis that the
Universe possesses cosmological CPT symmetry. According to this hypothesis,
all events in the Universe are symmetric with respect to the hypersurface
corresponding to the moment of cosmological collapse ($t=0$). He postulated
\cite{Sah2} this for time reflection $t\rightarrow -t$: $\psi \rightarrow
\gamma _{5}\psi $ ($C$-reflection for spinor fields) and $e_{\mu
a}\rightarrow -e_{\mu a}$ ($PT$-reflection of the tetrad field). In
\cite{Tur1,Tur2} a conformally flat metric $g_{\mu \nu }=\varphi
^{2}\left( t\right) \eta _{\mu \nu }$ near the singularity is considered,
where $\eta _{\mu \nu }$ is the Minkowski metric, and the conformal factor
is proportional to time: $\varphi (t)\sim t$. In this case, a $%
t\rightarrow -t$ isometry also arises, imposing an additional condition on
the vacuum of quantum field theory. This results in the emission of heavy
neutrinos from the singularity, similar to the Hawking radiation of black
holes. According to the authors, this provides the most elegant and
convincing model of dark matter currently available. However, this
formulation of the $CPT$ symmetry of the universe is valid only if the
spacetime geometry near the singularity remains classical. In quantum
gravity, the time-reflection operation $t\rightarrow -t$ remains undefined.
We know that the Wheeler-DeWitt equation (WDW) for the wave function of the
universe does not contain a time parameter. In Gibbons and Hawking's \cite%
{GH} formulation of Euclidean quantum gravity, this is expressed, as we
shall see, by an additional integration of the Feynman functional integral
over the time parameter. Thus, to formulate $CPT$ symmetry in quantum
gravity, we need a clearer understanding of the time parameter in this
theory.

In this paper, a time parameter of the universe that is invariant with
respect to transformations of $3D$ coordinates (diffeomorphisms) and gauge
transformations is determined. Dividing the symmetry group of the theory
into three components -- internal symmetries of matter fields, $3D$
diffeomorphisms and, actually, the dynamic semigroup -- allows us to
formulate the rules of $C$-, $P$- and $T$-transformations separately and the
$CPT$ symmetry of the mirror universe in the quantum theory of gravity
(QTG). The $CPT$ theorem states that any local relativistic quantum field
theory with a Lagrangian invariant under the Lorentz transformations is
necessarily invariant under the combined action of the $C$, $P$, and $T$
transformations. The possibility of transferring the $CPT$ theorem to
quantum cosmology is based on the structure of the canonical dynamics of a
closed universe in a certain class of reference systems similar to inertial
ones. Mirror reflection in a homogeneous isotropic model of a closed
universe is considered in detail.

In the next section, we define a class of equivalent reference frames in a
closed universe and a time parameter in each of them, for which the
operation of reflection about a singular point is defined. The operation of
space reflection is also defined in this construction. The third section
formulates a quantum theory of gauge fields in which charge conjugation is
explicitly taken into account. The fourth section introduces the definition
of a two-sheeted universe with $CPT$ mirror symmetry. A homogeneous
isotropic model of the universe with the mirror symmetry is considered in the
fifth section.

\section{PROPER TIME OF A CLOSED UNIVERSE}

In quantum mechanics, the reflection of time $t\rightarrow -t$ in the Schr%
\"{o}dinger equation is equivalent to the complex conjugation $\psi
\rightarrow \psi ^{\ast }$ of the wave function. We will retain this simple
rule in quantum gravitational theory if we find an appropriate definition of
time in a closed universe. We will use the operator representation of
gravitational constraints obtained in \cite{GL}. Based on this
representation, in \cite{GLG1} an operator analogue of the WDW equation is
also proposed, in which, together with the wave function, the time reference
system to which the quantum state of the universe belongs is also
determined. In this paper, the proper time of the universe and the
corresponding reference frame are introduced at the classical level, which
will allow to take into account the requirement of $T$-symmetry (as well as $%
C $- and $P$-symmetries) in the quantization procedure.

The operator representation is based on the Witten identity [9] in real
integral form obtained in \cite{GL}:

\begin{equation}
\left( \chi ,\widehat{W}\eta \right) =\int_{\Sigma }d^{3}x\left[ N\left(
\chi ,\eta \right) \widetilde{H}_{0}+N^{i}\left( \chi ,\eta \right)
\widetilde{H}_{i}\right] ,  \label{1}
\end{equation}%
where $\chi ,\eta $ are Dirac bi-spinors. Here $\widetilde{H}_{0},\widetilde{%
H}_{i}$ are the gravitational constraints in the real canonical
representation of Arnowitt, Deser and Misner \cite{ADM}, and the operator $%
\widehat{W}$ on the space of bi-spinors is defined as the difference of two
positive-definite, Hermitian operators (see \cite{GL}),

\begin{equation}
\widehat{W}\equiv \mathit{D}^{2}-\Delta .  \label{2}
\end{equation}%
Here

\begin{equation}
\mathit{D}\binom{\lambda ^{A}}{\mu ^{\ast A^{^{\prime }}}}=i\sqrt{2}\binom{%
n_{\left. {}\right. A^{^{\prime }}}^{A}\sigma _{\left. {}\right. \left.
{}\right. B^{^{\prime }}}^{\ast kA^{^{\prime }}}\nabla _{k}^{\ast }\mu
^{\ast B^{^{\prime }}}}{n_{\left. {}\right. A}^{A^{^{\prime }}}\sigma
_{\left. {}\right. B}^{kA}\nabla _{k}\lambda ^{B}}  \label{3}
\end{equation}
is $3D$ Dirac operator on the spatial section $\Sigma $, and $\Delta $ is
the sum of the Beltrami-Laplace operator with metric coefficients and the
energy-momentum tensor of the matter fields. $n_{AA^{^{\prime }}}$ is an
arbitrary unitary spin matrix, the simplest of which is

\begin{equation}
n_{AA^{^{\prime }}}=\frac{1}{\sqrt{2}}\left(
\begin{array}{cc}
1 & 0 \\
0 & 1%
\end{array}%
\right) ,  \label{4}
\end{equation}%
and $\sigma _{\left. AB\right. }^{k}$ are the spin coefficients of the $3D$
metric $\beta _{ik}$ on $\Sigma $ (see \cite{GL}). For the simplest choice
of $n_{AA^{^{\prime }}}$ in the representation Eq.(\ref{4}), the spin
coefficients of the metric are real: $\sigma _{\left. A^{^{\prime
}}B^{^{\prime }}\right. }^{\ast k}=\sigma _{\left. AB\right. }^{k}$. The
covariant derivative of the spinor field

\begin{equation}
\nabla _{k}\lambda ^{A}=\partial _{k}\lambda ^{A}+A_{k\left. {}\right.
\left. B\right. }^{\left. {}\right. A}\lambda ^{B}  \label{5}
\end{equation}%
here is constructed using Ashtekar's complex connection $\left( \text{\cite%
{GL}}\right) $,

\begin{equation}
A_{k\left. {}\right. \left. B\right. }^{\left. {}\right. A}=\Gamma _{k\left.
{}\right. \left. B\right. }^{\left. {}\right. A}\left( \sigma \right) +\frac{%
i}{\sqrt{2}}M_{k\left. {}\right. \left. B\right. }^{\left. {}\right. A},
\label{6}
\end{equation}%
in which $M_{k\left. {}\right. \left. B\right. }^{\left. {}\right. A}$ are
the canonical coordinates of the theory of gravity in the Ashtekar
representation, conjugate to the spin densities $\widetilde{\sigma }_{\left.
AB\right. }^{k}$, playing the role of canonical momenta, and $\Gamma
_{k\left. {}\right. \left. B\right. }^{\left. {}\right. A}\left( \sigma
\right) $ is a real spin connection, which uniquely determines the
Christoffel connection \cite{Fran}. In our case, the components of $%
M_{k\left. {}\right. \left. B\right. }^{\left. {}\right. A}$ are also real.
The indices $A,A^{^{\prime }}=0,1$ are raised and lowered using the
antisymmetric spin tensor $\varepsilon _{AB}=-\varepsilon _{BA}=1$. The
Hermitian scalar product of bi-spinors is defined as

\begin{equation}
\left( \chi ,\chi \right) =\int_{\Sigma }\sqrt{\beta }d^{3}xn_{AA^{^{\prime
}}}\left[ \lambda ^{\ast A^{^{\prime }}}\lambda ^{A}+\mu ^{A}\mu ^{\ast
A^{^{\prime }}}\right] ,  \label{7}
\end{equation}%
where $\beta =\det \beta _{ik}$. Finally, the coefficients on the right-hand
side of the Eq.(\ref{1}) (the lapse and shift functions \cite{MTW}) are
equal to:

\begin{equation}
N\left( \chi ,\eta \right) =\frac{1}{8}n_{AA^{^{\prime }}}\left[ \lambda
_{1}^{\ast A^{^{\prime }}}\lambda _{2}^{A}+\mu _{1}^{A}\mu _{2}^{\ast
A^{^{\prime }}}\right] ,  \label{8}
\end{equation}

\begin{equation}
N^{k}\left( \chi ,\eta \right) =\frac{i}{4}\sigma _{\left. AB\right. }^{k}%
\left[ \lambda _{1}^{A}\lambda _{2}^{+B}+\mu _{1}^{A}\mu _{2}^{+B}\right] ,
\label{9}
\end{equation}%
and

\begin{equation}
\lambda ^{+A}=\sqrt{2}n_{\left. {}\right. A^{^{\prime }}}^{A}\lambda ^{\ast
A^{^{\prime }}}  \label{10}
\end{equation}%
- involution operation ($\lambda ^{++A}=-\lambda ^{A}$).

Let us consider the secular equation for the operator $\widehat{W}$:

\begin{equation}
\widehat{W}\chi =W\chi .  \label{11}
\end{equation}%
The eigenvalue can be represented as:

\begin{equation}
W=\frac{\left( \chi ,\widehat{W}\chi \right) }{\left\vert \left\vert \chi
\right\vert \right\vert ^{2}},  \label{12}
\end{equation}%
where $\chi $ is the corresponding eigenvector. These quantities are
functionals of all fundamental canonical variables of gravity theory. We
will show that the eigenvalue $W$ is invariant under $3D$ coordinate
transformations on $\Sigma $. This is obvious already from the $3D$
covariance of the secular equation Eq.(\ref{11}). However, it is useful to
verify this by direct calculation. We introduce the normalized eigenvector $%
\widetilde{\chi }=\chi /\left\vert \left\vert \chi \right\vert \right\vert $
and calculate the Poissone bracket (PB) commutator

\begin{equation}
\delta W=\left\{ \int_{\Sigma }d^{3}x\zeta ^{k}\widetilde{H}_{k},\left(
\widetilde{\chi },\widehat{W}\widetilde{\chi }\right) \right\} ,  \label{13}
\end{equation}%
where $\zeta ^{k}\left( x\right) $ is the infinitesimal shift of spatial
coordinates. We will use identity Eq.(\ref{1}), in which we will assume that
$\widetilde{\chi }$ is canonically neutral. We will also take into account

\begin{eqnarray}
\delta \sigma _{\left. AB\right. }^{k}\left( y\right) &=&\left\{
\int_{\Sigma }d^{3}x\zeta ^{k}\widetilde{H}_{k},\sigma _{\left. AB\right.
}^{k}\left( y\right) \right\}  \notag \\
&=&-\sigma _{\left. AB\right. }^{l}\left( y\right) \nabla _{l}\zeta
^{k}\left( y\right) ,  \label{14}
\end{eqnarray}%
which is consistent with

\begin{equation}
\delta \beta _{ik}=-\nabla _{i}\zeta _{k}-\nabla _{k}\zeta _{i}.  \label{15}
\end{equation}%
Using the algebra of constraints \cite{Fran}, we find:

\begin{equation}
\delta W=-\int_{\Sigma }d^{3}x\left[ \delta N\left( \chi ,\chi \right)
\widetilde{H}_{0}+\delta N^{k}\left( \chi ,\chi \right) \widetilde{H}_{k}%
\right] ,  \label{16}
\end{equation}%
where

\begin{eqnarray}
\delta N\left( \chi ,\chi \right) &=&\frac{1}{8}n_{AA^{^{\prime }}}\left[
\delta \lambda ^{A}\lambda ^{\ast A^{^{\prime }}}+\lambda ^{A}\delta \lambda
^{\ast A^{^{\prime }}}\right.  \notag \\
&&\left. +\delta \mu ^{A}\mu ^{\ast A^{^{\prime }}}+\mu ^{A}\delta \mu
^{\ast A^{^{\prime }}}\right] ,  \label{17}
\end{eqnarray}

\begin{eqnarray}
\delta N^{k}\left( \chi ,\chi \right) &=&-\frac{i}{4}\sigma _{\left.
AB\right. }^{k}\left[ \delta \lambda ^{A}\lambda ^{+B}+\lambda ^{A}\delta
\lambda ^{+B}\right.  \notag \\
&&\left. +\delta \mu ^{A}\mu ^{+B}+\mu ^{A}\delta \mu ^{+B}\right] .
\label{18}
\end{eqnarray}%
In this case, for example,

\begin{equation}
\delta \lambda ^{A}=-\zeta ^{k}\nabla _{k}\lambda ^{A},  \label{19}
\end{equation}%
where is the covariant derivative of the spinor field $\lambda ^{A}$,

\begin{equation}
\nabla _{k}\lambda ^{A}=\partial _{k}\lambda ^{A}+\Gamma _{k\left. {}\right.
\left. B\right. }^{\left. {}\right. A}\left( \sigma \right) \lambda ^{B},
\label{20}
\end{equation}%
is defined using the spin connection $\Gamma _{k\left. {}\right. \left.
B\right. }^{\left. {}\right. A}\left( \sigma \right) $. However, it is easy
to verify that in formulas Eqs.(\ref{17}) and (\ref{18}) it can be extended to
the complex Ashtekar connection, $\Gamma _{k\left. {}\right. \left. B\right.
}^{\left. {}\right. A}\rightarrow A_{k\left. {}\right. \left. B\right.
}^{\left. {}\right. A}$. We need this extension of the connection in order
to again use Witten's identity Eq.(\ref{1}) and write:

\begin{equation}
\delta W=\left( \delta \widetilde{\chi },\widehat{W}\widetilde{\chi }\right)
+\left( \widetilde{\chi },\widehat{W}\delta \widetilde{\chi }\right) =0.
\label{21}
\end{equation}%
The equality to zero is explained by the fact that the eigenvectors $%
\widetilde{\chi }$ of the Hermitian operator $\widehat{W}$ form an
orthonormal set. Thus, the variation Eq.(\ref{13}) of the eigenvalue $W$
under infinitesimal spatial shifts is equal to zero (provided that $%
\widetilde{\chi }$ are canonically neutral). But relation Eq.(\ref{21})
itself proves the canonical neutrality of $\widetilde{\chi }$ as a
consequence of the orthonormality of the set of eigenvectors! The invariance
of the eigenvalues of the operator $\widehat{W}$ with respect to $3D$
diffeomorphisms is proven. The PB commutator of two eigenvalues $\left\{
W_{1},W_{2}\right\} $, which reduces to a linear combination of the
generators of the diffeomorphisms $\widetilde{H}_{k}$, can also be found by
direct calculation. We will not need this commutator further. Here we will
take the path of fixing spatial coordinates on $\Sigma $. Following the
Faddeev-Popov procedure \cite{FadPop}, we reduce the phase space by solving
the constraints $\widetilde{H}_{i}$ together with additional coordinate
conditions. This procedure does not violate the locality of the theory.

Then, we can replace the set of Hamiltonian constraints $\widetilde{H}_{0}$,
quadratic in the momenta (one for each point in space), by a set of
eigenvalues $W$, each of which is defined on the entire section $\Sigma $ in
the reference frame determined by the corresponding eigenvector $\widetilde{%
\chi }$. Thus, the \textquotedblleft multi-arrow\textquotedblright\ time
\cite{MTW} is replaced by a certain class of reference systems and
corresponding time parameters on $\Sigma $, determined by the spectrum of
the operator $\widehat{W}$. Since $\widetilde{\chi }$ are canonically
neutral, the canonical dynamics of GR in all these systems has the same
form. This class of reference frames is analogous to the class of inertial
reference frames in the special theory of relativity. Below, we will
quantize the GR in one of the reference frames of this class, considering
the initial action

\begin{eqnarray}
I_{GR}\left[ 0,T\right] &=&\int_{0}^{T}dt\left\{ \int_{\Sigma }d^{3}x\left[
\widetilde{\sigma }_{\left. AB\right. }^{k}\overset{\cdot }{M}_{k}^{\left.
{}\right. AB}+\widetilde{p}_{\alpha }\overset{\cdot }{\phi }_{\alpha }\right]
\right.  \notag \\
&&\left. -\left( \widetilde{\chi },\widehat{W}\widetilde{\chi }\right)
\right\} .  \label{22}
\end{eqnarray}

Due to the canonical neutrality of $\widetilde{\chi }$, the Hamiltonian
function and the entire action integral Eq.(\ref{22}) are local. The moment $%
t=0$ corresponds to the beginning of the expansion of the universe. Keeping
in mind the reflection of time relative to this point, we further extend the
interval under consideration to $[-T,T]$. Note that the connections $%
\widetilde{H}_{k}$ - generators of diffeomorphisms are not included in the
action Eq.(\ref{22}). The time reference system we fixed is determined by
the solution of the $3D$ covariant equation Eq.(\ref{11}) in an arbitrary $%
3D $ coordinate system on $\Sigma $. This ensures the $3D$ invariance of
action Eq.(\ref{22}). In fact, without loss of generality, we will solve the
secular equation Eq.(\ref{11}) in special spherical coordinates on the $3D$
sphere, in which arbitrary tensor fields of spin coefficients $\sigma
_{\left. AB\right. }^{k}$ and $M_{k}^{\left. {}\right. AB}$, as well as
tensor fields of matter $\phi _{\alpha }$ and $\widetilde{p}_{\alpha }$ are
given. Spherical coordinates on a $3D$ sphere have two poles - "north" and
"south," so spatial reflection ($P$---inversion) can be defined simply as a
change of poles on $\Sigma $. Thus, along with the reflection of time, in
the secular equation Eq.(\ref{11}), we perform a spatial reflection - a
change of poles on $\Sigma $. We will discuss charge conjugation ($C$%
-inversion) combined with the quantization procedure in the next section.

Before doing so, however, let us make a remark about the quantization
procedure for covariant theories. The wave function of the universe in the
QTG is independent of the time parameter $T$. As a consequence of the
Batalin, Fradkin, and Vilkovinsky (BFV) theorem \cite{BFV}, in the simplest
case of a relativistic particle, the invariant functional integral must
include additional integration over the parameter $T$ in the interval $\left[
0,\infty \right) $ with a simple measure $\mu (T)=1$ \cite{Gov}. For a
theory with quadratic momentum constraints (with a two-dimensional modular
group \cite{Gov}), forward motion in time occurs precisely at $T>0$. This
statement can be extended to the GR in a fixed reference frame $\widetilde{%
\chi }$, and we can immediately write down the covariant propagator of the
universe (without its mirror image yet):

\begin{equation}
\mathit{K}=\int_{0}^{T}dT\int \prod\limits_{t,x}\mu _{FP}D\sigma DMDpD\phi
\exp \left\{ \frac{i}{\hslash }I_{GR}\left[ 0,T\right] \right\} ,  \label{23}
\end{equation}%
where $\mu _{FP}$ is the Faddeev-Popov measure. The no-boundary wave
function of the Hartle-Hawking universe \cite{HH} is obtained from here by
passing to the Euclidean form of the functional integral under additional
smoothness conditions at the initial time $t=0$. Our observation is that
when the time interval in the action functional Eq.(\ref{22}) and the
corresponding time reflection rules are extended to $[-T,T]$, the outer
integral in Eq.(\ref{23}) remains unchanged.

The second remark concerns the behavior of the canonical time parameter
under the reflection $T\rightarrow -T$. The structure of the operator $%
\widehat{W}$ is such that the trace of the spin tensor $M_{k}^{\left.
{}\right. AB}$, $M_{k}^{\left. {}\right. AB}\sigma _{\left. AB\right. }^{k}$
is contained only in the Dirac operator Eq.(\ref{3}). It follows that if $d$
is some eigenvalue of the Dirac operator,

\begin{equation}
\mathit{D}\eta =d\eta ,  \label{24}
\end{equation}%
then it, being equal

\begin{equation}
d=\left( \widetilde{\eta },\mathit{D}\widetilde{\eta }\right) ,  \label{25}
\end{equation}%
where $\widetilde{\eta }=\eta /\left\Vert \eta \right\Vert $, is canonically
conjugate to the $3D$ volume of the spatial section $\Sigma $

\begin{equation}
V=\int_{\Sigma }\sqrt{\beta }d^{3}x,  \label{26}
\end{equation}%
which, in turn, PB- commutes with the rest of the $\Delta $ operator $%
\widehat{W}$. This means that the $3D$ volume is the true canonical time
parameter of the universe, and when reflecting $T\rightarrow -T$, the
reflection $\sqrt{\beta }\rightarrow -\sqrt{\beta }$ must also hold. We add
that the spectrum of eigenvalues of the $3D$ Dirac operator Eq.(\ref{3}) is
two-valued, and the eigenvalues themselves have the physical meaning of the
energy of a closed universe. Thus, the reflection of proper time in
cosmology is equivalent to a change in the sign of the universe's energy,
which again leads to a complete analogy with a relativistic particle.

\section{CHARGE CONJUGATION AND QUANTIZATION OF GAUGE FIELDS}

The action integral Eq.(\ref{20}) lacks a contribution corresponding to the
internal gauge symmetries of the matter fields. In conventional gauge
theory, the Lagrangian density in the canonical form of the action must be
supplemented with the constraints $\Phi _{\sigma }$ with the corresponding
Lagrangian multipliers: $\varphi _{\sigma }\Phi _{\sigma }$. The constraints
form a closed algebra with respect to the Poisson brackets:

\begin{equation}
\left\{ \Phi _{\sigma },\Phi _{\sigma ^{^{\prime }}}\right\} =C_{\sigma
\sigma ^{^{\prime }}\sigma ^{\prime \prime }}\Phi _{\sigma ^{^{\prime \prime
}}},  \label{27}
\end{equation}%
where $C_{\sigma \sigma ^{^{\prime }}\sigma ^{\prime \prime }}$ are the
structure constants. In the case of a compact semisimple Lie group $SU(2)$
\cite{FS}

\begin{equation}
\left\{ \Phi _{a}\left( x\right) ,\Phi _{b}\left( y\right) \right\}
=g\varepsilon _{abc}\Phi _{c}\left( x\right) \delta ^{3}\left( x-y\right) ,
\label{28}
\end{equation}%
where $g$ is the interaction constant, and the Lagrange multipliers $\varphi
_{a}$ coincide with the components $A_{0}^{a}$- the time component of the
4-vector of the Yang-Mills field $A_{\mu }^{a}$. In the case of an
electromagnetic field, $A_{0}$ is a scalar potential, the sign of which in
the presence of a point charge is determined by the sign of the charge.
Thus, the charge reflection operation ($C$-inversion) acts primarily on the
Lagrange multipliers $\varphi _{\sigma }$. At the same time, the Lagrange
multipliers $\varphi _{\sigma }$ have certain transformation properties with
respect to the infinitesimal gauge transformations generated by $\Phi
_{\sigma }$:

\begin{equation}
\delta \varphi _{\sigma }=\overset{\cdot }{\epsilon }_{\sigma }-C_{\sigma
\sigma ^{^{\prime }}\sigma ^{\prime \prime }}\varphi _{\sigma ^{^{\prime
}}}\epsilon _{\epsilon ^{^{\prime \prime }}},  \label{29}
\end{equation}%
which ensure the invariance of the action functional. We rewrite Eq.(\ref{29}%
) as a functional differential equation

\begin{equation}
\frac{\delta \varphi _{\sigma }\left( t\right) }{\delta \epsilon _{\sigma
^{^{\prime }}}\left( t^{^{\prime }}\right) }=\delta _{\sigma \sigma
^{^{\prime }}}\frac{d}{dt}\delta \left( t-t^{^{\prime }}\right) -C_{\sigma
\sigma ^{^{\prime \prime }}\sigma ^{\prime }}\varphi _{\sigma ^{^{\prime
\prime }}}\left( t\right) \left( t-t^{^{\prime }}\right)  \label{30}
\end{equation}%
with respect to the functional $\varphi _{\sigma }\left( t,\left[ \epsilon
_{\sigma ^{^{\prime }}}\left( t^{^{\prime }}\right) \right] \right) $,
supplementing it with the initial condition $\varphi _{\sigma }\left( 0,%
\left[ \epsilon _{\sigma ^{^{\prime }}}\left( t^{^{\prime }}\right) \right]
\right) =0$. It is easy to see that its solution has the form

\begin{equation}
\varphi _{\sigma }\left( t\right) =\overset{\cdot }{\varphi }_{\sigma
^{^{\prime }}}\left( t\right) \Lambda _{\sigma \sigma ^{^{\prime }}}\left(
\epsilon \left( t\right) \right) ,  \label{31}
\end{equation}%
where $\Lambda _{\sigma \sigma ^{^{\prime }}}\left( \epsilon \left( t\right)
\right) $ can be represented as a series:

\begin{eqnarray}
\Lambda _{\sigma \sigma ^{^{\prime }}}\left( \epsilon \right) &=&\delta
_{\sigma \sigma ^{^{\prime }}}-\frac{1}{2!}C_{\sigma \sigma ^{^{\prime
\prime }}\sigma ^{\prime }}\epsilon _{\sigma ^{^{\prime \prime }}}  \notag \\
&&+\frac{1}{3!}C_{\sigma \sigma ^{^{\prime \prime }}\rho ^{^{\prime
}}}C_{\rho ^{^{\prime }}\rho \sigma ^{\prime }}\epsilon _{\sigma ^{^{\prime
\prime }}}\epsilon _{\rho }-\cdot \cdot \cdot .  \label{32}
\end{eqnarray}%
We now write the action of gauge theory in the form:

\begin{equation}
I_{G}=\int_{0}^{T}dt\left[ p_{k}\overset{\cdot }{q}_{k}-h-\overset{\cdot }{%
\epsilon }_{\sigma ^{^{\prime }}}\Lambda _{\sigma \sigma ^{^{\prime
}}}\left( \epsilon \right) \Phi _{\sigma }\right] .  \label{33}
\end{equation}%
If we assume that under a finite gauge transformation $\epsilon (t)$ the
action Eq.(\ref{33}) does not change, we conclude that the dynamic variables
($q_{k},p_{k}$ ) now also have a finite gauge shift $\epsilon (t)$ relative
to the origin $t=0$. In other words, the additional dynamic variables $%
\epsilon (t)$ determine the position of the system on the manifold of the
gauge group, and the action Eq.(\ref{33}) describes the motion of the system
over time along this manifold. In fact, we now add the third term under the
integral sign Eq.(\ref{33}) to the GR action Eq.(\ref{22}). In quantum
theory, such an addition means that the Schr\"{o}dinger equation

\begin{equation}
i\hslash \frac{\partial \psi }{\partial t}=\left( \widetilde{\chi },\widehat{%
\widehat{W}}\widetilde{\chi }\right) \psi ,  \label{34}
\end{equation}%
describing the motion of the universe in time must be supplemented by the
Tomonaga-Schwinger equation,

\begin{equation}
i\hslash \frac{\delta \psi }{\delta \epsilon _{\sigma }\left( x\right) }%
=_{\sigma ^{^{\prime }}}\Lambda _{\sigma \sigma ^{^{\prime }}}\left(
\epsilon \right) \widehat{\Phi }_{\sigma }\psi ,  \label{35}
\end{equation}%
which describes the independent motion of the universe on a compact manifold
of the gauge group. Now the wave function of the universe depends not only
on time $t$, but also on the local gauge shift $\epsilon _{\sigma }\left(
x\right) $. We will write the solution of the system Eqs.(\ref{34}) and (\ref%
{35}) as before in the form of a functional integral for the propagator, and
we will not forget that in covariant quantum theory additional integration
over time is necessary, and now over all other gauge variables:

\begin{eqnarray}
\mathit{K} &=&\int_{0}^{T}dt\int_{G}\prod\limits_{t,\Sigma }D\epsilon \int
\prod\limits_{t,\Sigma }\mu _{FP}D\sigma DMDpD\phi  \notag \\
&&\times \exp \left\{ \frac{i}{\hslash }I_{GR}\left[ 0,T;0,\epsilon _{\sigma
}\left( x\right) \right] \right\} .  \label{36}
\end{eqnarray}%
The second outer integral is taken over the gauge group manifold at each
point $\Sigma $. This is the covariant propagator of the universe
originating at the Big Bang singularity, where $t=0$ and $\epsilon _{\sigma
}\left( x\right) =0$. The no-boundary wave function of the Hartle-Hawking
universe is obtained from this by transitioning to the Euclidean form of the
functional integral with additional smoothness conditions at the origin. Our
task now is to perform a $CPT$ reflection about the origin and formulate the
principle of $CPT$ symmetry in quantum cosmology.

\section{CPT SYMMETRY IN QUANTUM COSMOLOGY}

In order to speak of symmetry in reflection relative to the universe's
origin, it is necessary to have this same universe "on both sides" of the
singularity. Therefore, we will now write down the action of a two-sheeted
universe over an extended time interval:

\begin{eqnarray}
&&I_{GR}\left[ T,-T;\epsilon _{\sigma }\left( x\right) ,-\epsilon _{\sigma
}\left( x\right) \right]  \notag \\
&=&\int_{-T}^{T}dt\left\{ \int_{\Sigma }d^{3}x\left[ \widetilde{\sigma }%
_{\left. AB\right. }^{k}\overset{\cdot }{M}_{k}^{\left. {}\right. AB}+%
\widetilde{p}_{\alpha }\overset{\cdot }{\phi }_{\alpha }\right. \right.
\notag \\
&&\left. \left. -\overset{\cdot }{\epsilon }_{\sigma ^{^{\prime }}}\Lambda
_{\sigma \sigma ^{^{\prime }}}\left( \epsilon \right) \Phi _{\sigma }\right]
-\left( \widetilde{\chi },\widehat{W}\widetilde{\chi }\right) \right\} ,
\label{37}
\end{eqnarray}%
where the integration is performed along an arbitrary curve in the extended
(due to $\epsilon _{\sigma }$ $(x,t)$) phase space between two fixed
boundary points, excluding additional conditions that denote the transition
between the universe and its mirror image. The conditions are as follows: i)
the time coordinate $T$ and the group coordinates $\epsilon _{\sigma }$ $(x)$
of the boundary points have opposite signs, ii) $\sqrt{\beta }=\det \sigma
_{k}^{AB}>0$ $\left( <0\right) $ when $t\in \left( 0,T\right] $ $\left( t\in %
\left[ -T,0\right) \right) $, iii) when $t\rightarrow -t$ the poles of the
spherical spatial coordinates change to $\Sigma $ (spatial reflection).
Condition ii) means that the singular point of the universe divides it into
two mirror sheets with opposite signs of the scale factor. Given these
additional conditions, the covariant propagator of a two-sheeted universe is:

\begin{equation}
\mathit{K}_{mir}=\int_{0}^{T}dt\int_{G}\prod\limits_{t,\Sigma }D\epsilon
\mathit{K}\left[ T,-T;\epsilon _{\sigma }\left( x\right) ,-\epsilon _{\sigma
}\left( x\right) \right] ,  \label{38}
\end{equation}%
where

\begin{eqnarray}
&&\mathit{K}\left[ T,-T;\epsilon _{\sigma }\left( x\right) ,-\epsilon
_{\sigma }\left( x\right) \right]  \notag \\
&=&\int \prod\limits_{t,\Sigma }\mu _{FP}D\sigma DMDpD\phi  \notag \\
&&\times \exp \left\{ \frac{i}{\hslash }I_{GR}\left[ T,-T;\epsilon _{\sigma
}\left( x\right) ,-\epsilon _{\sigma }\left( x\right) \right] \right\} .
\label{39}
\end{eqnarray}%
Note that the limits of the outer integration with respect to the time
parameter have not changed: propagator Eq.(\ref{38}) still describes forward
motion in time. But now the origin of the motion is the mirror state of the
universe in the past, and the singularity is located at the midpoint.

Let us now formulate the principle of $CPT$ symmetry in the mirror universe,
for which we should define its initial state $\psi _{in}\left[ -T,-\epsilon
_{\sigma }\left( x\right) \right] $ in the equation

\begin{eqnarray}
\psi _{out}\left[ T,\epsilon _{\sigma }\right] &=&\int \prod\limits_{\Sigma
}d^{6}Md^{SM}\phi \mathit{K}\left[ T,-T;\right.  \notag \\
&&\left. \epsilon _{\sigma },-\epsilon _{\sigma }\right] \psi _{in}\left[
-T,-\epsilon _{\sigma }\right] .  \label{40}
\end{eqnarray}%
According to the definition in ordinary quantum mechanics, it should
obviously be:

\begin{equation}
\psi _{in}\left[ -T,-\epsilon _{\sigma }\left( x\right) \right] =\psi
_{out}^{\ast }\left[ -T,-\epsilon _{\sigma }\left( x\right) \right] .
\label{41}
\end{equation}%
The rules of spatial reflections (pole reversals) and the sign reversal of $%
\det \sigma _{k}^{AB}$ are also preserved. The latter is performed
automatically upon pole reversal. Equation Eq.(\ref{41}) clearly uniquely
determines the state of the universe at any moment in time $T$ and at any
point in the gauge group. The final step in our construction of a covariant
quantum theory is to calculate the outer integrals over these parameters in
Eq.(\ref{38}). However, we will take one more step, which is necessary near
the singularity: we will move to the Euclidean form of the theory.

The transition to the Euclidean functional integral using the Wick rotation
in the complex time plane $T\rightarrow iT$ is necessary to determine the
propagator in Eq.(\ref{39}). Before this, we perform a formal functional
integration over all canonical momenta, which yields an infinite
normalization factor \cite{FS}. We obtain:

\begin{eqnarray}
&&\mathit{K}_{E}\left[ T,-T;\epsilon _{\sigma }\left( x\right) ,-\epsilon
_{\sigma }\left( x\right) \right]  \notag \\
&=&\int \prod\limits_{t,\Sigma }d^{6}Md^{SM}\phi  \notag \\
&&\times \exp \left\{ -\frac{1}{\hslash }I_{GR}^{L}\left[ T,-T;\epsilon
_{\sigma }\left( x\right) ,-\epsilon _{\sigma }\left( x\right) \right]
\right\} ,  \label{42}
\end{eqnarray}%
where $I_{GR}^{L}$ is the Lagrangian form of the action Eq.(\ref{37}). Now
we formulate the principle of $CPT$ symmetry in the Euclidean form of the
theory. We can still use relation Eq.(\ref{40}) (with the replacement $%
\mathit{K}\rightarrow \mathit{K}_{E}$), but instead of Eq.(\ref{41}) we
write:

\begin{equation}
\psi _{in}\left[ -T,-\epsilon _{\sigma }\left( x\right) \right] =\psi _{out}%
\left[ -T,-\epsilon _{\sigma }\left( x\right) \right] ,  \label{43}
\end{equation}

\section{HOMOGENEOUS ISOTROPIC MIRROR UNIVERSE}

Euclidean action of Einstein's theory of gravity,

\begin{equation}
I_{HE}=-\frac{1}{16\pi G_{N}}\int \sqrt{g}d^{4}xR+I_{m}\left[ g,\phi \right]
,  \label{44}
\end{equation}%
where $I_{m}\left[ g,\phi \right] $ is the action of the (standard model)
matter fields, in the case of a homogeneous isotropic model of the universe
with a metric

\begin{equation}
ds^{2}=N^{2}\left( \tau \right) d\tau ^{2}-a^{2}\left( \tau \right)
dS_{3}^{2},  \label{45}
\end{equation}%
$dS_{3}^{2}$ is the metric interval of a $3D$ sphere of unit radius and one
real scalar field $\phi $, and takes the form:

\begin{eqnarray}
I_{FL}\left[ a,\phi \right] &=&\int_{0}^{1}d\tau N\left\{ -\frac{a}{2\kappa }%
\left[ \left( \frac{\overset{\cdot }{a}}{N}\right) ^{2}+1\right] \right.
\notag \\
&&\left. +2\pi ^{2}a^{3}\frac{1}{2}\left[ \left( \frac{\overset{\cdot }{\phi
}}{N}\right) ^{2}+m^{2}\phi ^{2}\right] \right\} ,  \label{46}
\end{eqnarray}%
where $\kappa =2G_{N}/3\pi $. Action Eq.(\ref{46}) is invariant with respect
to arbitrary transformations of the parameter $\tau $ (reparametrizations) $%
\tau ^{^{\prime }}=\tau ^{^{\prime }}\left( \tau \right) $ provided that

\begin{equation}
N=\overset{\cdot }{c},  \label{47}
\end{equation}%
where $c\left( \tau \right) $ is a scalar function of the parameter $\tau $.
This quantity is in this case the proper time of the universe. For the no
boundary wave function of Hartle and Hawking, this case is considered in
detail in \cite{HHH1,HHH2}. We can immediately write the
functional-integral representation of the no boundary wave function of the
universe in Euclidean form:

\begin{eqnarray}
\psi _{NB}\left( b,\chi \right) &=&\int_{0}^{\infty }dC\int DaDp_{a}D\phi
Dp_{\phi }  \notag \\
&&\times \exp \left\{ -\frac{1}{\hslash }\int_{0}^{C}dc\left[ p_{a}\overset{%
\cdot }{a}+p_{\phi }\overset{\cdot }{\phi }-\mathit{H}_{FL}\right] \right\} ,
\notag \\
&&  \label{48}
\end{eqnarray}

\begin{equation}
\mathit{H}_{FL}=-\frac{1}{2}\left( \frac{\kappa p_{a}^{2}}{a}-\frac{a}{%
\kappa }\right) +\frac{1}{2}\left( \frac{p_{\phi }^{2}}{2\pi ^{2}a^{3}}-2\pi
^{2}a^{3}m^{2}\phi ^{2}\right) .  \label{49}
\end{equation}%
Here, integration over proper time (with a simple measure $\mu \left(
C\right) =1$) is limited by positive values of the two-valued modular group
of reparametrization invariance \cite{Gov}, and the Feynman functional
integral is written in the canonical representation \cite{FS}. Integration
in the configuration space is carried out along trajectories $\left( a\left(
c\right) ,\phi \left( c\right) \right) $ with boundary values

\begin{equation}
a\left( 0\right) =0,\overset{\cdot }{\phi }\left( 0\right) =0,  \label{50}
\end{equation}

\begin{equation}
a\left( C\right) =b,\phi \left( C\right) =\chi .  \label{51}
\end{equation}%
Strictly speaking, integration over a should be restricted to positive
values. The boundary conditions at the origin are called smoothness
conditions. The functional integral Eq.(\ref{48}) can be estimated using the
saddle-point method, which yields \cite{HHH1,HHH2}:

\begin{equation}
\psi _{NB}\symbol{126}\sum \exp \left\{ -\frac{1}{\hslash }I_{FL}\left[
\widetilde{a},\widetilde{\phi }\right] \right\} .  \label{52}
\end{equation}%
Here the summation is carried out over all extremal Euclidean trajectories $%
\left( \widetilde{a}\left( c\right) ,\widetilde{\phi }\left( c\right)
\right) ,c\in \left[ 0,C\right] $ satisfying the boundary conditions Eqs.(%
\ref{44}) and (\ref{45}), where $C$ is determined by the Hamiltonian constraint
equation in Euclidean form,

\begin{equation}
\mathit{H}_{FL}\left[ b,\chi ,\hslash \frac{\partial }{\partial b},\hslash
\frac{\partial }{\partial \chi }\right] \psi _{NB}=0,  \label{53}
\end{equation}%
as a consequence of the estimate of the integral over $C$ in Eq.(\ref{48}).
Equation (\ref{53}) coincides with the WDW equation for the wave function
$\psi _{NB}\left( b,\chi \right) $. Note that extremal Euclidean
trajectories form complex conjugate pairs, so the wave function $\psi
_{NB}\left( b,\chi \right) $ is real. We recalled the structure of the
ordinary univalent universe with no boundary wave function.

Now let's move on to a two-sheeted universe with a mirror image. The
starting point will be the Euclidean propagator (42). In this case, it is
equal to:

\begin{eqnarray}
\mathit{K}_{FL}\left[ C,-C\right] &=&\int DaDp_{a}D\phi Dp_{\phi }  \notag \\
&&\times \exp \left\{ -\frac{1}{\hslash }\int_{-C}^{C}dc\left[ p_{a}\overset{%
\cdot }{a}+p_{\phi }\overset{\cdot }{\phi }-\mathit{H}_{FL}\right] \right\} .
\notag \\
&&  \label{54}
\end{eqnarray}%
The singularity $a(0)=0$ is now in the middle of the integration domain, and
on the interval $[-C,0]$, the functional integration is carried out over
negative values of $a$. When calculating the integral Eq.(\ref{54}) using
the saddle-point method, in addition to Eqs.(\ref{50}) and (\ref{51}), the
boundary conditions will be needed

\begin{equation}
a\left( -C\right) =b^{^{\prime }},\phi \left( -C\right) =\chi ^{^{\prime }}.
\label{55}
\end{equation}%
The conditions in the middle of the classical trajectory are preserved,
including the smoothness condition $\overset{\cdot }{\phi }\left( 0\right)
=0 $. This is required by the classical equation of motion of a scalar field

\begin{equation}
\overset{\centerdot \centerdot }{\phi }+3\frac{\overset{\centerdot }{a}}{a}%
\overset{\centerdot }{\phi }-m^{2}\phi =0  \label{56}
\end{equation}%
to ensure smoothness at $a(0)=0$. However, in contrast to the single-sheet
model of the universe \cite{HHH1,HHH2}, the value of $\phi \left(
0\right) $ is not an arbitrary parameter of the problem, but is determined
by the boundary conditions Eqs.(\ref{51}) and (\ref{55}).

Now we define the initial state of the two-sheeted universe using the mirror
symmetry condition Eq.(\ref{37}). Here we have no gauge parameters $\epsilon
_{\sigma }$, and the initial wave function $\psi _{in}$ is equal to the
final wave function $\psi _{out}$ if we make the substitutions $C\rightarrow
-C,b\rightarrow -b^{^{\prime }}$ in the latter. We will take the covariance
condition (the integral over $C$) into account at the very end of the
constructions. Instead of explicitly solving the problem of finding the
mirror initial condition, we determine it iteratively. As a zero
approximation, we take $\psi _{0in}\left( -C,b^{^{\prime }},\chi ^{^{\prime
}}\right) =\psi _{NB}\left( -C,-b^{^{\prime }},\chi ^{^{\prime }}\right) $.
In the first approximation, we obtain:

\begin{eqnarray}
\psi _{1out}\left( C,b,\chi \right) &=&\int_{-\infty }^{0}dC\int_{-\infty
}^{+\infty }d\chi ^{^{\prime }}\mathit{K}_{FL}\left[ C,-C;b,\chi
;b^{^{\prime }},\chi ^{^{\prime }}\right]  \notag \\
&&\times \psi _{0in}\left( -C,b^{^{\prime }},\chi ^{^{\prime }}\right) .
\label{57}
\end{eqnarray}%
In the second approximation:

\begin{eqnarray}
\psi _{2out}\left( C,b,\chi \right) &=&\int_{-\infty }^{0}dC\int_{-\infty
}^{+\infty }d\chi ^{^{\prime }}\mathit{K}_{FL}\left[ C,-C;b,\chi
;b^{^{\prime }},\chi ^{^{\prime }}\right]  \notag \\
&&\times \psi _{1in}\left( -C,b^{^{\prime }},\chi ^{^{\prime }}\right) .
\label{58}
\end{eqnarray}%
And so on. One might expect that after a large number of iterations the
state of the universe will \textquotedblleft forget\textquotedblright\ the
initial no boundary wave function and it will turn out that

\begin{equation}
\psi _{Nin}\left( -C,b^{^{\prime }},\chi ^{^{\prime }}\right) \simeq \psi
_{Nout}\left( -C,-b^{^{\prime }},\chi ^{^{\prime }}\right) .  \label{59}
\end{equation}%
In this case, we can speak of a multi-sheeted, oscillatory model of the
universe. As a final step, this stable oscillatory state should be
integrated over $C$:

\begin{equation}
\psi _{N}\left( b,\chi \right) =\int_{0}^{\infty }dC\psi _{Nout}\left(
C,b,\chi \right) .  \label{60}
\end{equation}%
Here it is necessary to pay attention that the iterative wave function $\psi
_{Nout}\left( C,b,\chi \right) $, in addition to the initial dependence on $%
C $ in the boundary of the wave function $\psi _{NB}\left( -C,-b^{^{\prime
}},\chi ^{^{\prime }}\right) $ and the final dependence on $C$, contains
these dependencies in intermediate links of the form $\mathit{K}_{FL}\left[
C,-C\right] \mathit{K}_{FL}\left[ -C,C\right] $. If we evaluate integral Eq.(%
\ref{60}) using the saddle-point method in the semiclassical approximation,
all intermediate contributions to the extremum condition with respect to $C$
cancel each other out. As a result, we are left with the extremum condition
for a finite dependence on $C$ $\psi _{Nout}\left( C,b,\chi \right) $. Thus,
the wave function $\psi _{N}\left( b,\chi \right) $, as well as $\psi
_{NB}\left( b,\chi \right) $, defined by Eq.(\ref{53}), is, in the indicated
approximation, a solution to the WDW equation Eq.(\ref{53}). This solution,
although constructed as a two-sheeted one, describes a multi-sheeted
oscillating universe.

\section{CONCLUSIONS}

Formally, the two-sheeted model of the universe proposed in this paper is an
alternative solution to the WDW equation, which differs from the no-boundary
Hartle-Hawking solution in its boundary conditions near the singularity. The
requirement of $CPT$ symmetry replaces the smoothness conditions of the
Euclidean functional integral at the origin. However, the physical content
of the two-sheeted solution is significantly different: instead of the birth
of the universe from \textquotedblleft nothing,\textquotedblright\ we can
talk about the singularity as a physical phenomenon that obeys certain
symmetry laws, namely $CPT$ symmetry. This event serves as a boundary (or
link) between two epochs of quantum evolution of the universe with a $CPT$%
-symmetric material substrate. The proposed theory allows us to return to
the problem of the baryon asymmetry of the universe within the framework of
QTG near a cosmological singularity. Moreover, the procedure of quantization
of gauge fields of matter, which includes in explicit form the operation of
charge conjugation, may have significance outside the cosmological issues
considered here.

\section{ACKNOWLEDGEMENTS}

We are thanks V.A. Franke for useful discussions.




\end{document}